# Exploring Information Centrality for Intrusion Detection in Large Networks

*Nidhi Rastogi, Rensselaer Polytechnic Institute*

## Abstract

Modern networked systems are constantly under threat from systemic attacks. There has been a massive upsurge in the number of devices connected to a network as well as the associated traffic volume. This has intensified the need to better understand all possible attack vectors during system design and implementation. Further, it has increased the need to mine large data sets, analyzing which has become a daunting task. It is critical to scale monitoring infrastructures to match this need, but a difficult goal for the small and medium organization. Hence, there is a need to propose novel approaches that address the big data problem in security. Information Centrality (IC) labels network nodes with better vantage points for detecting network-based anomalies as central nodes and uses them for detecting a category of attacks called systemic attacks. The main idea is that since these central nodes already see a lot of information flowing through the network, they are in a good position to detect anomalies before other nodes. This research first dives into the importance of using graphs in understanding the topology and information flow. We then introduce the usage of information centrality, a centrality based index, to reduce data collection in existing communication networks. Using IC-identified central nodes can accelerate outlier detection when armed with a suitable anomaly detection technique. We also come up with a more efficient way to compute Information centrality for large networks. Finally, we demonstrate that central nodes detect anomalous behavior much faster than other non-central nodes, given the anomalous behavior is systemic in nature.

## Introduction

From a cyber security perspective, the year 2017 was challenging for organizations all over the world. Businesses of all sizes have been impacted in unprecedented ways. In a survey, Kaspersky labs gathered data from over 4000 companies, big and small, across the world about the most crucial metric of incident detection and response, which is time. Reportedly, 8.2% were able to detect a security event almost instantly, 14.19% within a few hours, and 20.5% within a day of the attack. These numbers indicate a growing need for security intelligence for businesses. The deep-rooted cause behind such alarming statistics is the rapid deployment of connected devices. IDS are burdened with analyzing massive datasets and hence require more time to perform their function. The report is proof of how detection time has increased and that even the most advanced IDSs are unable to cater to the changing demands of the security domain. A study [2] performed by Cisco suggests that the data related situation will exacerbate as the annual global IP traffic will increase to 2.3 zeta bytes by the year 2020. This will increase the challenges in separating useful data from noise, and identifying attack vectors from the datasets. This is one of the causes that delay attack detection. It is also the main motivation for this research. In this research, we identify volume anomalies (or systemic anomalies), a class of anomaly that has widespread consequence. They are beneficial in detecting systemic attacks that result in the failure of an entire system as opposed to an individual component. Some of the potential outcomes of a systemic cyber-attack are denial of service, significant delay in meeting client requests, and at times cascading effects on the supporting backbone infrastructure [37]. Distributed denial of service (DDoS) attacks have been identified as the top threat for any Internet facing entity [20]. In a DDoS, multiple compromised machines attack a target and render certain services inaccessible for the legitimate user of those services. We make the assumption that the platform itself is trustworthy and initially uncompromised. It is also imperative to have a well-defined understanding of the kinds of attacks or threats that the system needs protection from. Same detection principle cannot apply to all kinds of attacks. A spear phishing attack targets one clearly identified user and has a different attack signature. This is in contrast to a distributed denial of service (DDoS) attack where many compromised systems attack a single target system. An attack signature is a unique way of identifying a pattern of information that describes an attempt to exploit a system. Specifically, we consider network layer attacks that impact an entire system. These attacks are sometimes also referred as systemic attacks. An attacker can launch them through various means, like :

a) malware where a malicious program gets installed on a vulnerable machine and spreads throughout the network by infecting other vulnerable machines in close contact,
b) DDoS by sending incessant network requests to several devices thereby preventing useful packets to reach the desired destinations in the network. This attack, called denial-of-service, is usually distributed in nature and originates from zombie machines remotely controlled by the attacker machine(s), or
c) steal large amount of data from one or more end nodes, etc.

We propose a novel way for network-wide anomaly detection. It sparsifies overall data collection and analyses by accumulating it from the influential nodes in the network. Consequentially, this increases the chances of attack detection and in a timely fashion. This approach further reduces computational needs for anomaly detection and ultimately decreases analysis time. We also back our theory, through simulations and experimentation on network datasets, that it is possible to identify cyber threats from a smaller network data, even without compromising the capability of the system to detect attacks. Complex systems like communication networks can often be understood by bottom-up analysis. Investigating the behavior of individual components and their interconnections can reveal interesting insights into the system as a whole. Interactions between components and other entities can be studied using graphs where nodes are connected to each other over links known as edges. Data gets transferred between entities over these links. We use existing, graph based techniques called centrality measures to identify important nodes in the network. These special nodes, also known as central nodes, are important because information propagates through them the fastest, the most, or has the maximum reach. In our context, we use centrality algorithm that identifies nodes which are central to information propagation in the network. To prove our hypothesis of using IC for network-wide anomaly detection, we do the following:

a) Model a simulation based network and analyze its behaviors in the event of system-wide anomalies. Earlier work [4] shows that simulation based experimentation is able to explore diverse approaches to solving problems in network security. Simulations are able to do this as they can manage tradeoffs in a complex system, where conducting real world experiments can be difficult or may require a very specific network dataset.
b) Our approach recommends the process of identifying and usage of specific nodes in a network where defenses can be deployed.

Essentially, we examine the impact of identifying central nodes in deploying defenses so that they can defend against any kind of network-wide attack. We also answer a few related questions.

*a) How does topology factors into effectively identifying these nodes for deploying defense mechanisms?*
*b) What role does communication frequency play along with the topology in identifying central nodes?*

For this, we use a graph-based approach from social network studies which has never been used for this purpose to the best of our knowledge. Our key contributions are as follows:

a) Introduce a novel approach for minimizing data for analysis in large networks. To this end, we analyze data traversing through the most important nodes in the network, known as central nodes; and
b) Through simulation based experiments, we prove that the time taken to uncover anomalous behavior in the network drastically reduces anomaly detection time, when central nodes used for analysis.

The remainder of the paper is divided into the following sections: First we cover important concepts used in this research. Then, we delve into the main research idea and its implementation details. The simulation environment that is used to demonstrate the implementation is also covered. Finally, we discuss the results of the simulation and analyze them.

## Literature Study

Large networks send enormous amount of data from one device to another in order to exchange information. For this, they select routes that add the least overhead to the network, and yet can be used to reliably communicate with other nodes. In these networks, aside from the scale related issues due to overwhelming number of devices and their connectivity, understanding the behavior of the network at a given time is very challenging and poses a series of obstacles. The success of a network monitor is therefore crucially dependent on how timely can it report metrics that are useful in understanding the behavior of the network, including any noteworthy changes. These changes can direct the network analyst towards points of issues impacting the network and its users. Contemporary network monitors use intrusion detection systems, firewalls, network and device event logs, or a combination of these. Besides straightforward techniques that measure deviation from the threshold of network measurement attributes, there are other statistical techniques that have been successfully deployed. Network-based intrusion detection systems (NIDS) protect IT systems and resources from external miscreants that steal information, render systems partially or completely inoperable, etc. However, existing N-IDS techniques are unable to capture typical network conditions to recognize disruptions, and only rely on fixed thresholds to detect those events that cause extreme changes. Such events necessitate continuous monitoring and learning the usual routing dynamics and thereafter, deviations in the behavior of the network before deciding to investigate an event. Also, with systemic attacks, limited investigation was performed as the focus was on the router level of networks. We build on these earlier approaches, but considerably extend them by enabling the analysis that benefits a large class of anomaly detection. The drawback of tracking these behaviors continuously is the enormous amount of data that they generate. N-IDS can be broadly classified into two categories:

    a) Misuse detection - Signatures created for past malicious behaviors are cataloged for future usage.
    b) Anomaly detection, where behavior outside of modeled baseline behavior is flagged and appropriate action taken to contain it.

### *Anomaly Detection*

In this section, we provide an overview of the research in this field. Anomaly detection is a well-researched and a high-impact application that spans across varied research areas and domains [5]. Both generic as well as application-specific anomaly detection techniques have been developed over the course of past two decades. By definition, anomaly detection refers to finding patterns in data that do not conform to expected behavior of the system. It often results in critical, actionable information in a wide variety of application domains. For example, if an anomalous traffic pattern is observed in a computer network, it could mean that a hacked computer is sending out sensitive data to an unauthorized machine. The process of anomaly detection can be summarized by two methods – firstly, what needs to be detected, and secondly – how detection takes place. These two phases of identifying intrusion seek distinctive information from the collected data. While misuse detection searches for description that matches a known malicious behavior, anomaly detection works on the notion of separating normal behavior from the rest. This allows anomaly detection to recognize attacks that never been seen before (also known as zero-day attacks).

## Machine Learning based Anomaly Detection

Before anomaly detection can be applied to a system or dataset, the exact notion of an anomaly needs to be understood as it can be different for different application domains. For instance, in healthcare domain a small deviation from the normal range of TSH level might be an anomaly, while similar deviation in the insurance business might be considered as normal. Thus, anomaly detection cannot be applied using the similar approach among different domains to another, is not straightforward. It actually depends on the following criteria:

a) Availability of labeled data for training or validation of models
b) Ease of distinguishing and separating useful data from noise.

Chandola et. al [5] extensively cover various anomaly detectors as well as the application domain and knowledge disciplines they are developed for. In [1], the author suggested that a key requirement for any technique to work, one needs a sound understanding of the system. We are of the opinion too that the most critical requirement for creating a relevant tool for any environment is to gain deep insights into the system and its functioning, capabilities, and limitations. Treating one like a black box and expecting the detection tool to work instantly, will lead to erroneous results and very high false-positives. A commonly deployed fix is to increase the sensitivity of the tool. But one cannot guarantee this will remain an effective strategy over time. Due to these challenges, the anomaly detection problem is not easy to solve. Although, machine learning offers a wide range of tools for today's systems, such as neural networks, support vector machines, etc., most of the existing anomaly detection techniques solve a specific formulation of the problem. For this research, we propose a graph-based approach which addresses the issue of creating customized solutions for each category of anomalous behaviors in a given network. Graphs can be used to model a wide variety of structures and relationships, for example networks, and offer commonly used terminologies to describe them. Since graphs have the ability to analyze a system from both a top-down and a bottom-up approach, they offer ways to formulate problems and also possible approaches to solve them. They also provide tools to investigate systems and to study their behavior given a set of certain input conditions. A number of research efforts have used graphs analytics result, which are produced by network flow records to detect malicious traffic. [38] uses graph cut for intrusion detection [42] is designed to detect anomaly when the graph structure changes. The paper [7] is based on commute time, which tracks abnormal changes in graph structure and edge weight due change in node relationships. The published research in ML based anomaly detection so far has a few shortcomings. Their primary focus is not on communication network and it does not appear in their research. Copious research work addresses the advantages of sparsifying or condensing graph datasets, but none uses it for anomaly detection.

## Graph Analytics based Anomaly Detection

Akoglu et. al [13] extensively cover various outlier detection schemes that are based on graphs. The authors recommend graph based anomaly detection for several cases, especially when the ground truth is not available or is already marked with anomalies. Graph-based approach is also useful when the data is multidimensional and cannot be reduced to fewer ones. In communication networks classification models can be trained for link-based spam detection [17]. It takes advantage of features like average degree of neighbor, PageRank [19], TrustRank [16], etc. Online spam filtering [18] on social networks is another area which uses graph principles like sender's degree, interaction history, incremental clustering, etc. In the field of computer networks, network intrusion uses graph based network feature representations to analyze network traffic [15]. This is the closest to which our research is, but varies in the approach and overall detection goals. This motivated us into exploring graph-based anomaly detection. The main advantages of these methods are:

a) Network graphs can be explored for their correlations over long periods of time [13],
b) Graphs are not analyzed in isolation, but in correlation to other entities,
c) Multi-dimensional characteristics can be analyzed at a singular level as well as in clusters, both big and small.

## *Systemic Attack identification using Anomaly Detection*

The focus of our anomaly detector is to defend against systemic attacks such as DDoS. Literature provides several detection mechanisms; however, each defense mechanism has been designed to accommodate different aspects of the attack problem. We summarize a few of them below:

a) Detection point - The attack can be diagnosed either on the network or on the host machines like servers, etc. The network detects attacks when the packets route through network junctions, such as gateways etc. On the other hand, individual hosts detect potential attacks targeted at them. We model only network based attacks.

b) Detection principle - An attack can be detected either by using an existing attack signatures, a network layer congestion pattern, protocol behavior, or host-based activities. Their efficacy lies in identifying a certain type of attack pattern. In our simulations, we characterize the detection criteria by using an existing machine learning technique, which identifies likelihood of false positives and false negatives.

This research focuses on proposing a detection mechanism that can be automatically triggered at network nodes in the presence of an on-going network-wide attack.

## Centrality measures and Networks

We now we go over the importance of using centrality measures in networks. The concept of centrality indices, or just centrality, originally came in late 1940s from studying human communication in small groups [7]. Soon centrality gained a strong foothold as an estimate of an individual's importance in social networks [8]. It established relationships between certain features of a network and an actor's influence. In other words, it identified "the most important actor" or "the most useful entity" in a social graph (where a vertex is an actor), alibi using ad-hoc formalization.

To understand better, consider a directed acyclic graph (DAG), G(V, E) where V represents the vertices and E, the edges connecting them. These vertices represent nodes in a network and the direction of the link defines the direction of data flow. This graph is called a network graph and represents the link structure of nodes in the network. Since an edge corresponds to a link over which data packets are transferred from a node, this embodies the idea that the link contains relevant information. Simultaneously, a node plays an important role in transferring this data to its final destination by acting as an intermediary. And, the node is able to do that precisely because of its location and connection to the destination node. Thus, it is reasonable to assume that the structural relevance of a node with respect to other nodes in the graph is called the centrality of the node.

With time, many centrality measures were introduced that were motivated by the idea that an individual with proximity to others will have more information [9], have more power [11], greater prestige [12], or have a greater influence [21] than others. These measures and linkages between actors were used in several network analytic studies to evaluate fairly large networks successfully [29], however only for application domains like social networks, community organizations, and planning for the most part. Despite many useful properties, centrality measures suffer from a major drawback. When not used for the appropriate network, or type of network flow, they can often lead to incorrect understanding of results. If a centrality index does not relate to the purported index, it becomes difficult to understand the measurement. Hence, a clear understanding of the concept of centrality type and its potential for an application is very important. In the next section, we summarize the different flow processes based on which the choice for centrality measure depends.

## Understanding Centralities and Flow Processes

In this section, we cover some of the most widely applied centrality measures in the field of social networks. There are different ways of measuring centrality. However, whether a metric is applicable to a particular network depends on the two dimensions of typology - the trajectories followed by network flows, and spreading mechanism. Network traffic may take the shortest path (geodesic), a path (do not repeat nodes or links), a trail (do not repeat links) or a walk (can repeat nodes and links). The method of spread can be broadcasting, parallel duplication, serial duplication, or transfer. Data is copied to each node with both parallel and serial duplication but not with transfer. In [10], the author shows through simulations that these metrics are independent of the underlying graph or network structure. Along the lines of [10], we review some of the well-known centrality measures in order to understand the flow assumption they make.

a) Closeness centrality measures for a node, the sum of all graph-theoretic distances from all other nodes in a connected graph. Nodes with low scores for closeness centrality will tend to receive flows sooner. Here the assumptions are that flows originate from all nodes with equal probability, and along the shortest path. Closeness is applicable for situations where traffic travels along all routes, like package delivery. And therefore, it can be interpreted as an index of arrival time of the traffic. It however allows parallel transfer of information across several routes.

b) Betweenness centrality counts the number of shortest paths that pass through a node. In essence, it measures the total control of a node on the volume of traffic that flows through that node. This measure will be applicable for traffic that is transferred from node to node along the shortest route, instead of being copied or broadcasted. The traffic has a target and knows the best ways to get there. Diseases or gossip do not follow this pattern, as they are copied and not moved. They also do not search for a specific target(s), which even if is the case, may not travel along the shortest path.

c) Eigenvector centrality measures whether a node has neighbors who themselves score high on centrality. It counts the number of walks of all lengths, weighted inversely by length, which emanate from a node. In essence, traffic moves unrestricted along any route, where each node follows a parallel duplication process in which it affects all of its neighbors simultaneously. This measure is suited well for processes that measure influence.

d) Degree centrality measures the total number of links incident upon a node. Since it measures a kind of flow process that is independent to the typology applicable to closeness, betweenness, and eigenvector. It involves only direct links and hence measures the immediate influence. Also, all parallel duplication flow processes apply to this measure of centrality. As noted, none of these centrality measures are suitable for transferring data in a communication network. The one that comes close is the closeness centrality, which does not measure active information transfer along the shortest path. If the path exists, it is considered in measuring the centrality. Information centrality considers information flow along all paths. Since it is related to closeness (information is the reciprocal of path length), it can also be interpreted as an index of arrival time of the traffic. This methodology is used to calculate the central nodes in the simulation network.

e) Information centrality is founded on the axioms of centrality measure - an index based ranking system in the field of graph analytics, which orders important nodes from most to least central. According to this algorithm, these central nodes have the potential to control information flow in the network.

We argue that the behavior of these critical nodes can be tapped to check the pulse of the entire network. Since only these specific nodes will be monitored, it automatically reduces the amount of data analyzed, computational complexity, and time to detect an anomalous occurrence in the network. Since emphasis is on analyzing fewer nodes and lesser associated data, it is a crucial aspect of our research idea.

We demonstrate it by running a simulation based experiment on the network simulation platform, NS2. We further evaluate the performance of the algorithm by comparing the time taken to detect anomalies by central versus non-central nodes. Our results show that a significant improvement in the time required to detect anomalous behavior in a network when central nodes are used. Although, this approach is optimized for deployment on controlled nodes with limited energy and memory capacity, it can be extended to other networks. We use a simplified network with tractable features to contain the associated complexity, and yet understand the impact of our approach.

This work is informed by recent conceptual work that identifies the suitability of centrality indices in communication networks [30]. It exhibits a close relationship between flow of current in an electrical network and random walks around a graph. We use these results to motivate our search for a centrality measure applicable to networks, and ultimately find central nodes to assist in anomaly detection. We explore information centrality (IC) as a method to identify central nodes from a given network. It uses the location and connectivity of nodes in a given graph and identifies the most accessible ones. The central nodes are quicker to respond or to identify any systemic change in the network behavior because of this property. Because of this primary reason, we explore IC to detect anomalous behavior that impacts an entire system.

## Notations and Definitions

Let G = (V, E) be a connected, directed graph where V denotes the vertex set of size n and E denotes the edge set. Two vertices, i and j are considered adjacent if they lie at the two ends of an edge, E (i, j). If there is a vertex k in G connecting E(i,j) and E(j,k), then these edges are incident to each other on the vertex j. We assign weights to these edges, which is denoted by $wi,j$. The directed distance d(i,j) between i and j is considered to be the length of the directed shortest path from i to j. It is straightforward to realize that d(i,j) does not necessarily equals d(j,i), i.e. d(i, j) is not a metric of distance but a pseudo-metric [31]. A path is a sequence of vertices (v1,v2,...,vn), where each vertex is connected to the following vertex in the sequence. A path from a source vertex i to a destination vertex j is denoted as Pij = (i,i1,i2,...,iz-1,j). Clearly, there can be m multiple paths from i to j each denoted by Pij(r), where r = 1,..., m, r ε m and is the r th path from i to j (in some fixed ordering). Note that Pij (r) and Pij (r+1) are not necessarily disjoint. For vertices i,j, the weighted length [22], denoted by len of path Pij (r) is:

$$len(P_{i,j}(r)) = \sum_{e \in P_{i,j}(r)} \frac{1}{w(e)}$$

## Threat Model

We now describe our formal model comprising the network and related traffic, attacker profile, attack characteristics, and detection mechanism.

### Network

We model a network of n interconnected units with sensors that accumulate information by observing data shared among individual sensing units. Depending upon the placement of these units in the network and the traffic pattern, some may prove to be more influential and reliable in the system's decision-making process, especially in the presence of noisy data [25]. Accurate identification of these influential sensors may prove crucial to understanding the behavior of unusual occurrences in the network. Information centrality assigns to each node a quantity that, as we demonstrate in the next section, reflects its influence in the network. It identifies the structural elements of the network as well as the communication frequency between nodes to derive this number. We abstract this network of n units with sensors as a graph G with n vertices in V as defined in section 5. These vertices are connected point-to-point using private and authenticated channels [23]. There is a central adversary who could either be an individual entity or an organization with unbounded computing power, financial, and technological resources.

### Network Traffic

Body The traffic flow comprises a set of packets that are identified via a given set of traffic features (e.g., source and destination IP addresses and ports, and protocol). In this work, in order to study the evolution of flow, time is divided into fixed sized intervals. The volume of flow is the number of bytes in the flow during the corresponding interval. A flow in a given source node is generated by a discrete-time marked point process [26] and is independent of flow properties belonging to other nodes. The traffic is normal when the flow process remains the same over time. We also assume that the network links do not get saturated even in the presence of high volumes of anomalous traffic. The network behavior is considered abnormal when the flow process drastically increases due to an increase in the number of packets or bytes in the flow. In our paper, this is as a result of an attack by the adversary.

# Adversary and Attack

In this section, we show how the attack model is justified from the viewpoint of an adversary. We capture attacks using standard cryptographic terminology. To actively corrupt a unit, the attacker takes advantage of an existing vulnerability in the network units and uses an existing malignant code to exploit it. This code allows full control over the infected unit after which it displays arbitrary behavior. The adversary is assumed to be static and starts the attack process by corrupting t ϵ V units [27]. The attacker's eventual goal is to disrupt the functioning of the system and render it inaccessible to legitimate users. For this, the attacker may implement a distributed denial of service attack using the t nodes and use a broadcasting protocol to flood the network with excessive traffic or ping requests. As described in [24], a node v ϵ V can be in either of the two possible states infected or safe. At the beginning of the attack protocol, all nodes are safe and engaged in normal communication activities with the rest of the network. The attacker starts the protocol by infecting t nodes and changing their state from safe to infected. These infected nodes communicate with other safe nodes and spread the infection along the path of the network routing protocol. Once infected, a node is set to infected state and does not switch back to the safe state. The infected nodes behavior is controlled by the malicious code installed on them. The attack propagates along the links from any of the t infected nodes to the rest of the safe nodes that communicate with them and by replacing the code that exists in safe nodes. The attacker is oblivious to the change of state of infected nodes and cannot read the state or the behavior demonstrated by these nodes. The adversary (or attacker) actively corrupts networked units. The attacker neither compromises the integrity and authenticity of the communication nor reveals any cryptographic secret. The malicious code in every node gives more volume to the network by sending a gibberish payload to other units. Concretely, in this research we consider only volume related adversarial impact, one that has systemic impact on a network. We detect volume anomalies by observing link traffic of a large network. By definition, volume anomalies impact a major part of the network and refer to a sudden change in the network traffic [25]. This anomaly can originate either from outside or inside the network and is thus observed on most of the links traversed. In this work, we only consider internal nodes as the origin of the anomaly. There can be multiple causes leading to such anomalies, for example, a distributed denial of service attack causing incessant ping requests on a port, or stealing large amount of data from various databases and servers, that are also part of the network.

# Detection

We deploy a machine learning based anomaly detector and uses it across all nodes in the networks. The role of the detector is not to perform a network wide anomaly detection, but to keep track of the timestamp when the volume changes across flows. While a volume change does not signify anomalous behavior, the machine learning based detector does so on the basis of past information and how the detector has been built. In the best case, we achieve the same detection accuracy as the underlying detection mechanism. For simplicity, we assume that it is 100% and is left as an open problem for future exploration.

## *Information Centrality-based Anomaly Detector*

In this section, we describe the notion of information Centrality (IC) and explain its relation to two features of a node - it's location [9], and frequency of traffic passing through it. To understand IC, we take a detour to first understand the motivating concept and introduce some useful related terminologies. Information centrality identifies influential or important nodes that are central to information propagation in a graph network. The concept of information transmission is originally inspired from resistance distance measurement in an electrical circuit [9],[28].

Now we define the network distance between two vertices, i and j. A fixed electrical resistor is placed on i and j with a battery connected across them. For simplicity, we consider resistors of 1 Ohm. The effective resistance R is calculated using the Kirchhoff and Ohm laws. However, the resistance distance is not calculated simply as the sum of resistances along the path connecting the two vertices. Resistance decreases as the number of routes increase. Consequentially, the total resistance, R is the sum of the reciprocals of individual paths that connect i and j.

Similarly, in a network, if a packet sent from the source node i, individual bytes may traverse through several paths to reach destination node j. The time taken by this byte sized information depends on the distance of each path. In graph networks, resistance is referred as information measure, a term used to denote the total information transferred between a pair of nodes.

Let us demonstrate the concept using an example. We have a graph with seven vertices — i, j, k, l, m, n, and o and eight edges — (i,j), (i,k), (m,j), (i,m), (i,l), (l,k), (k,o), and (n,l). The edges are incident. The information measure of i, Iij with all other vertices; i.e. i1 ,i2 , ..., Iin. Let's take the case of vertex i. For the pair of vertices i, j, there are three paths, Pij (1) = i-j, Pij (2) = i-k-j, Pij (3) = i-m-n-j. The information measure Iij is:

$$I_{ij} = \frac{1}{len(Pij\,(1)) + Pij\,(2) + Pij\,(3))}$$

If Ii refers to IC of vertex i, then it will be defined as the harmonic average of information associated with the path from i to all other nodes. The IC of Iii is defined as infinity. The general formula for information centrality for a node i is given below:

$$\frac{1}{Ii} = (\frac{1}{n} \sum_{j}^{n} \frac{1}{Iij})$$

Harmonic mean is chosen over simple average in this case because information measure is given in ratios. Harmonic mean transforms all values with different denominators, into those with same ones. It takes care of extreme values and thus, gives a more accurate average value that the simple mean [31].

## Discussion on IC

Arguably, information centrality is an approach, which is intuitive and the claims made with respect to expected results, predictable. In essence, a simple observation of nodes that handle a lot of traffic tells us that they are central to traffic propagation and should be considered influential. Consequentially, they are also well positioned to detect anomalies that have a systemic impact on the network. It may appear that the work is trivial in its goals and does not warranty further research. And that an analytical model should suffice in terms of understanding the applicability of IC for networks. This is indeed the case with the IC approach, but partially. There is more to it than meets the eye. The topological and dynamical properties of modern communication networks pose challenges making it hard to analyze them in-depth, theoretically. Routing schemes further introduce complexities making it hard to establish that information centrality is a sound approach for anomaly detection. Consequentially, in this paper, we simulate a network that emulates real networks and their traffic patterns. The simulation also preserves most of the interesting features previously observed in other research work but is simplistic enough for analysis. Our contribution in this research is that information centrality effectively identifies these nodes in modern networks and is a good indicator of systemic anomalies, and prove this experimentally in the next section.

## Experiment

So far, IC has remained under explored in the research community, and it's applicability and exact usage mechanisms have remained vague, especially with respect to communication networks. For the first time, we demonstrate the procedure of using IC through simulation-based experiments in a wireless network. We follow the theoretical analysis performed in earlier sections to establish the proper usage of IC in a large data network. We also demonstrate the role played by node location and frequency of communication between node pairs in identifying those that influence information flow. Once central nodes have been identified, we validate our hypothesis that they can detect anomalous behavior that impacts a network at a systemic level, earlier than non-central nodes. We consider IC measure over other centrality measures for two main reasons. Firstly, no other centrality algorithm incorporates communication or data propagation property along with a node's topological properties. And secondly, it was recently proven that IC can be applied to data networks, alibi no work has been done in this direction, to the best of our knowledge. In this section, we set the goals for this experiment, the setup environment, and the tools used to demonstrate our approach. We the analyze the results in section 9. The goal of this experiment is to show that IC-AD can identify central nodes in a wireless network and those nodes can inform of an anomalous occurrence in the network early on. We choose the network parameters carefully, and keeping in mind the extensibility and inter-operability of protocols and features in a real network environment and with other protocols. The experimental setup was run on a 64-bit MacOS enabled 13inch MacBook Pro. The processor is a 2.9 GHz Intel Core i7 and the memory is 8GB with 1600MHz DDR3. The MacOS was running on the OS X El Capitan, version 10.11.16. We use NS-2 (described in next section) to simulate the network, traffic, and apply our approach for anomaly detection. The post simulation trace file is analyzed using python and Java code and we use jupyter and eclipse to run these codes, respectively.

## Network Simulator-2 (NS-2)

NS-2 is an open source event-driven simulator widely used in the research community to simulate communication networks [32]. It is a time-based, discrete event driven simulator. It can be used to simulate a wide array of applications that use protocols like TCP, UDP, FTP, and many more. This enables NS-2 to model and simulate real networks like wireless sensor networks and the nodes or devices using the networks. The simulation code can be written with a lot of precision to demonstrate which event happened, at what time, and how. Data transfer between entities can be shown visually using the visualization tool called NAM. It is also popular as it can run on various platforms like UNIX, Mac and windows platforms and the code is transferrable among platforms. NS-2 can also be used to test security of the network by testing different kinds of network attacks like denial of service, hello flood attack, sinkhole attacks.

## Simulation

We substantiate our hypothesis and simulate a mid-sized ZigBee network. The network comprises 200 nodes over a mesh topology using ns2[33]. This setup allows us to corroborate our hypothesis in a simple environment with tractable networking components and behaviors. At the same time, the simulation parameters have been chosen such that there are common features with other popular wireless network technologies like Wi-Fi and wlan. IC-AD will work in any environment, so long information is exchanged among networking entities. Table 1 summarizes the general simulation parameters.

### Table 1: Simulation Parameters

| Parameter | Value |
|---|---|
| Number of nodes | 200 |
| Simulation Area | 100x100 |
| Total simulation time | 900 s |
| Number of simulations | 100 |
| Traffic Type | CBR |
| Queue Model | Queue/DropTail/PriQueue |
| Max. packets in Queue | 1000 |
| Packet rate | 2/ms |
| Packet Size | 0.5KB |
| Total connections | 35 |

We simulate a ZigBee network because it is a low cost, low power consuming, and a short-range wireless communication technology. Developed for wireless personal area network (WPAN) in early 2000, it is widely used in building automation control, and monitoring of IoT. The simulation network has a mesh topology as it allows multi-hop peer-to-peer communication, using the AODV routing protocol [39], and decentralized routing. The antenna model is omnidirectional, to enable communication from all directions.

These simulation parameters are chosen keeping in mind their conceptual extensibility into other popular, large-scale networks like Wi-Fi and WLAN. This setup allows us to corroborate our hypothesis in a simple environment with tractable networking components and behaviors. For example, AODV as a routing protocol allows multi-hop packet transfer across the network.

AODV protocol is based on the DSDV proactive algorithm [40]. The idea of DSDV, in turn, was based on Bellman Ford shortest path algorithm. It calculates the shortest path in a network, like Dijkstra, but is slower that its peer. The reason being, Bell-Ford considers negative weights in a network edge. The AODV routing algorithms maintains a routing table to keep track of every node and the closest neighbor hop that leads to it. This can also be pictured in a static network where the given node is as the root node and the routing table as the minimum spanning tree of the mesh network. Each routing table for a node is also periodically updated in case a node goes offline or out of range. This algorithm for AODV can be easily shown as an extension of a graph-based routing protocol. It is also extensible to other, more complex routing protocols used in wireless and wired networks and hence was the protocol of choice for our simulation.

Similarly, a mesh topology is generic and random and does not restrict us from extending this work into other topologies. We keep the nodes as static, which prevents routes from changing frequently. However, the future work of this research allows some movement in the network nodes. IC-AD is applied to a synthetic data set since the real data sets only provide anecdotal evidence of anomalies. The experiments using real datasets are in an initial stage and are not covered in this paper. Also, we do not compare the results of IC-AD with another anomaly detection model yet, as the goal is to mainly demonstrate its anomaly detection capability in a homogeneous network. We perform experiments to evaluate the following characteristics of IC-AD:

        a) IC can be used to analyze large networks,
        b) Simulations result in same top central nodes as calculated using IC-AD, and
        c) Central nodes detect anomalies in the network before non-central nodes.

## Creating Ground Truth

This is the training phase where the simulation begins by initializing normal packet transfer across different the network using constant bit rate (CBR). The node locations and the source-destination pairs in ns-2 are randomly generated using a python script. To simulate the ground truth, we initialize the network by transmitting an average traffic load of 0.5 Mbps from various source-destination node pairs. This process continues for 80 seconds. At the end of this timeframe, IC-AD is able to categorize central nodes and then use them to detect anomalies. The simulation is run 100 times for various combinations of source and destination nodes, thereby simulating the flow of data packets in the network and ensuring that we get a similar set of nodes as central nodes. The topology remains the same for all the simulations. Since IC is calculated based on the flow of information of an entire network, it can also be used for identifying anomalous behaviors. We experiment with higher traffic of about 10Mbps and 50Mbps (this is high for a ZigBee network) to emulate excessively high average traffic load. Trace files for 900 seconds were captured under these traffic rates. This has a systemic impact on the network and is counted as a deviation from the normal network behavior. We monitor the performance of all central and non-central nodes and keep track of the time when the anomalous behavior is detected w.r.t the anomaly injection time.

## Classification of Central Nodes

IC-AD calculates the average packet arrival-time for all nodes, both central and non-central in the network. By recording the time it takes for a packet to reach a node, we identify nodes that are reachable in the shortest time making them central in a network. For analysis, the top 15% (30) and 20% (40) nodes ranked in ascending order of average arrival time are selected as central nodes. These are not absolute values and are purely used to demonstrate the value of central nodes.

## Anomaly Detection

We then compare the percentage of central versus non-central nodes that have identified the anomaly since introduction into the network. For detecting the abnormal increase in network traffic rate, we use an unsupervised machine learning technique called, support vector machine (SVM) as described in [41]. It uses feature space for anomaly detection, which is application specific, and hence information needs to be captured accordingly. In our experiment, we analyze one dataset comprising of average traffic load. Each entry is a sequence of all the possible traffic rates requested by various nodes. In this research, the anomaly detector used by the nodes is assumed to be efficient and in this paper, we do not delve deep into the error rate of the detection mechanism.

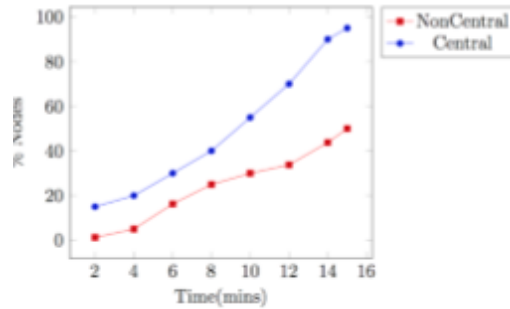

**Figure 1: Time taken to detect anomaly average traffic load = 10Mbps, central nodes = 15%**

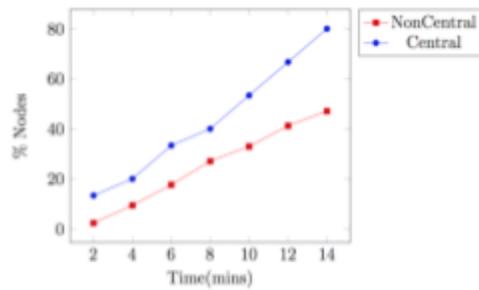

**Figure 2: Time taken to detect anomaly average traffic load = 10Mbps, central nodes = 20%**

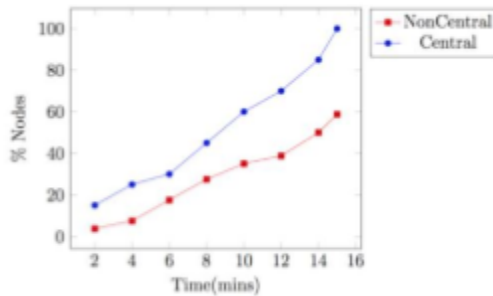

**Figure 3: Time taken to detect anomaly average traffic load = 10Mbps, central nodes = 15%**

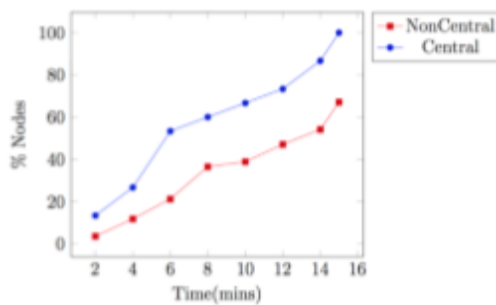

**Figure 4: Time taken to detect anomaly average traffic load = 10Mbps, central nodes = 20%**

## *Results and Analysis*

To evaluate IC-AD, we are interested in two major indicators of performance: the increasing order of average arrival time of data for nodes matches that of the order resulting from the information centrality algorithm. The graphs in Figures 1,2,3, and 4 show that our calculations are spot on target. The other indicator is that for the average traffic payload, the monitoring central nodes can detect the anomalous traffic rate much faster than the all other non-central nodes combined. As the anomaly propagates into the network the central nodes are able to detect it faster than the rest of the nodes. Monitoring the central nodes is a useful implementation in comparison to monitoring the entire network, as 90% or above central nodes become aware of the anomaly while only approximately 50%-60% (maximum case) of noncentral nodes for the duration of the simulation. We change the anomaly size and also decrease the central nodes monitoring the network to further boost our proposition. Results in the charts below show that the percentage of central nodes at any given point of time, always surpasses the total percentage of noncentral nodes that have detected the anomaly.

## *Summary*

Our main contributions in this research include proving that IC-AD can be used in communication networks for identifying central nodes. These central nodes are important and influence the flow of information, which makes them extremely valuable. IC-AD identifies these nodes based on their location and connectivity to other nodes using a harmonic average of the topological distance between nodes, and that central nodes can be used to study important behavioral changes that have a systemic impact on the network, such as anomaly detection. In comparison, other IDS use data coming from all the nodes in the network. Further, the simulation experiment lent confidence that IC-AD can analyze a network using fewer nodes and that property can be extended for anomaly detection.